\documentclass[aps,prd,twocolumn,showpacs]{revtex4}

\usepackage{amssymb}
\usepackage{graphicx}

\newcommand{\md}{{\mathrm{d}}}

\begin{document}

\title{Classical and quantum behavior of the harmonic and the quartic oscillators}

\author{David Brizuela}
\email{david.brizuela@ehu.es}
\affiliation{Fisika Teorikoa eta Zientziaren Historia Saila, UPV/EHU, 644 P.K., 48080 Bilbao, Spain}
\affiliation{Institut f\"ur Theoretische Physik, Universit\"at zu K\"oln, Z\"ulpicher Stra{\ss}e 77, 50937 K\"oln, Germany}

\begin{abstract}
In a previous paper a formalism to analyze the dynamical evolution of
classical and quantum probability distributions in terms of their moments was presented.
Here the application of this formalism to the system of a particle moving on a potential is considered
in order to derive physical implications about the classical limit of a quantum system.
The complete set of harmonic potentials
is considered, which includes the particle under a uniform force, as well as the harmonic
and the inverse harmonic oscillators. In addition, as an example of anharmonic system,
the pure quartic oscillator is analyzed. Classical and quantum moments corresponding
to stationary states of these systems are analytically obtained without solving any differential equation.
Finally, dynamical states are also considered in order to study the differences between
their classical and quantum evolution.
\end{abstract}


\maketitle

\section{Introduction}

Even if the foundations of the theory of quantum mechanics are very well settled, there are
still open questions about its classical limit
and the interaction between classical and quantum degrees of freedom.
In fact, there are hybrid theories which take into account classical
as well as quantum degrees of freedom (see for instance \cite{BoTr88, And95, ShSu78, PeTe01, Elz12, CHS12, BCG12}),
but will not be considered here. Concerning the classical limit of quantum mechanics,
in Ref. \cite{BYZ94} the idea that such a limit should be an ensemble of
classical orbits was proposed. This classical ensemble should be described
by a classical probability distribution on phase space and, thus, its evolution would be
given by the Liouville equation. It is not possible to compare directly classical and quantum
probability distributions since they are defined on different spaces. Therefore, a very convenient
way to perform such a comparison is by decomposing both probability distributions
in terms of its infinite set of moments. These moments are the observable quantities and one could
directly relate (and experimentally measure) their classical and quantum values.

The formalism to analyze the evolution of these moments was first developed in \cite{BM98}
for the Hamiltonian of a particle on a potential.
A formalism similar to this one, but with a different ordering of the basic variables,
was presented in \cite{BoSk06, BSS09} on a canonical framework
and for generic Hamiltonians.
Let us comment that this latter formalism has found several applications in the context of
quantum cosmology \cite{Boj12}. For example, isotropic models
with a cosmological constant have been analyzed \cite{BoTa08, BBH11}. Bounce scenarios
have also been studied within the framework of loop quantum cosmology \cite{Boj07}.
In addition, the problem of time has also been considered in \cite{BHT11, HKT12}.
Remarkably this framework is also useful when the dynamics is generated by a Hamiltonian constraint,
as opposed to a Hamiltonian function \cite{BoTs09}.

Recently the classical counterpart of the formalism developed in \cite{BoSk06, BSS09}
was presented \cite{Bri14}. In this reference it was argued that the quantum effects have
two different origins. On the one hand, \emph{distributional} effects are due to the fact
that, because of the Heisenberg uncertainty principle, one needs to consider an extensive
(as opposed to a Dirac delta) distribution with nonvanishing moments. These effects
are also present in the evolution of a classical ensemble and, for instance, they generically prevent the
centroid of the distribution (the expectation value of the position and momentum)
from following a classical trajectory on the phase space. On the other hand,
\emph{noncommutativity} or \emph{purely quantum} effects appear as explicit
$\hbar$ terms in the quantum equations of motion and have no classical counterpart.
In the present paper, this formalism for the evolution of classical and quantum probability
distributions will be applied to the case of a particle moving on a potential with the particular
aim of measuring the relative relevance of each of the mentioned effects.

The analysis will be made in two parts. On the one hand, the systems with a harmonic Hamiltonian
will be considered, that is, those that are at most quadratic on the basic variables.
This includes the system of a particle under a uniform force (which trivially includes
also the free particle case), the harmonic oscillator, and the inverse harmonic oscillator.
One of the properties of this kind of Hamiltonians is that there is no purely quantum effect
and, thus, they generate the same dynamics in the quantum and in the classical (distributional) cases.
In addition, the equations of motion generated by this harmonic Hamiltonians are much simpler
than in the general case, so it will be possible to obtain analytically the explicit form of their
moments corresponding to stationary as well as to dynamical states.

On the other hand, due to the complexity of the anharmonic case, a concrete particular example
must be analyzed. In our case, between the large set of anharmonic systems, we have chosen the pure quartic
potential in order to study both its stationary and dynamical states with
this formalism. As simple as it might seem, the quartic harmonic oscillator can not
be solved analytically and one usually resorts either to numerical
or analytical methods of approximation. Nonetheless, from a perturbative
perspective the model of the quartic oscillator corresponds to a singular
perturbation problem due to the fact that in the limit of a vanishing coupling constant,
several physical quantities diverge \cite{BW69, SiDi70}.
Hence, even if it has been studied during decades and, for instance,
its energy eigenvalues are well known
from numerical computations \cite{BOW77, Vor94}, this model is still considered of
interest in different context and new approximation techniques are
being developed to treat it, see e.g. \cite{LMT06, OBP12}.

The rest of the paper is organized as follows. In Sec. \ref{sec_formalism}
a summary of the formalism presented in Ref. \cite{Bri14} is given.
Section \ref{sec_potential} presents the equations of motion for a Hamiltonian
of a particle on a potential. In Sec. \ref{sec_harmonic} the harmonic cases
are analyzed. Section \ref{sec_anharmonic} deals with the anharmonic example
of the pure quartic oscillator. Finally, Sec. \ref{sec_conclusions} summarizes the
main results and details the conclusions of the paper.

\section{General formalism}\label{sec_formalism}

Given a quantum system with one degree of freedom described by the basic conjugate variables
$(\hat q, \hat p)$, it is possible to define the quantum moments as follows:
\begin{equation}
G^{a,b}:=\langle(\hat p - p)^a \,(\hat q - q)^b\rangle_{\rm Weyl}.
\end{equation}
In this equation $p:=\langle\hat p\rangle$ and $q:=\langle\hat q\rangle$ have been defined,
and Weyl (totally symmetric) ordering has been chosen.
The sum between the two indices of a given moment, $(a+b)$, will be referred as
its order.

The evolution equations for these moments are given by the following effective 
Hamiltonian, which is defined as the expectation value of the Hamiltonian operator
$\hat H$, and it is Taylor expanded around the position of its centroid $(q,p)$:
\begin{eqnarray} \label{HQ}
H_Q(q,p,G^{a,b})&:=&\langle\hat{H}(\hat q, \hat p)\rangle_{\rm Weyl}
\nonumber\\\nonumber
&=&\langle\hat{H}(\hat q-q+q, \hat p -p +p)\rangle_{\rm Weyl}\\\nonumber
&=& \sum_{a=0}^\infty\sum_{b=0}^\infty \frac{1}{a!b!}
\frac{\partial^{a+b} H}{\partial p^a\partial q^b} G^{a,b}\\
&=&H(q,p) + \sum_{a+b\geq 2}\frac{1}{a!b!}\frac{\partial^{a+b} H}{\partial p^a\partial q^b} G^{a,b}.
\end{eqnarray}
The Hamiltonian $H(q,p)$ is the function obtained by replacing in the Hamiltonian
operator $\hat H(\hat q, \hat p)$ every operator by its expectation value.

The equations of motion for the expectation values ($q$, $p$) and for the infinite
set of moments $G^{a,b}$ are directly obtained by computing the Poisson brackets between
the different variables with the Hamiltonian (\ref{HQ}). In particular, it is
easy to show that Poisson brackets between expectation values and moments vanish.
Furthermore, a closed formula is known for the Poisson bracket between any two moments \cite{BSS09,BBH11}.
In this way an infinite system of ordinary differential equations is obtained, which is completely
equivalent to the Schr\"odinger flow of states. In the general case, as will be shown
below, in order to perform the resolution of this system, it is necessary to introduce
a cutoff $N_{\rm max}$ and drop all moments of an order higher than $N_{\rm max}$.

The classical counterpart of this formalism is obtained by assuming a classical ensemble
described by a probability distribution function $\rho(\tilde q, \tilde p,t)$ on a
phase space coordinatized by $(\tilde q, \tilde p)$. As it is well known, the evolution
equation of such a distribution is given by the Liouville equation. Following the same
procedure as in the quantum case, making use of the probability distribution $\rho(\tilde q, \tilde p,t)$,
one can define a classical expectation value operation on the phase space:
\begin{equation}
\langle f(\tilde q,\tilde p) \rangle_{\rm c} := \int d\tilde q d\tilde p f(\tilde q,\tilde p) \rho(\tilde q,\tilde p,t),
\end{equation}
where the integration extends to the whole domain of the probability distribution.
With this operation at hand, the classical moments can be defined as
\begin{equation}
C^{a,b}:=\langle (\tilde p-p)^a (\tilde q-q)^b \rangle_{\rm c},
\end{equation}
$q$ and $p$ being the position of the centroid of the distribution, that is, $q:=\langle\tilde q \rangle_c$
and $p:=\langle\tilde p \rangle_c$.
Note that in this classical case, everything commutes and, thus, the ordering in the definition
of the moments is indifferent. As in the quantum case, the effective Hamiltonian that encodes the dynamical information
of these variables is constructed by computing the expectation value of the Hamiltonian and expanding it
around the position of the centroid. In this way, one obtains the classical effective Hamiltonian:
 \begin{eqnarray}
H_{\rm C}(q,p,C^{a,b})&:=&\langle  H(\tilde q,\tilde p) \rangle_{\rm c}\\\nonumber
&=&H(q,p) + \sum_{a+b\geq2}\frac{1}{a!b!}\frac{\partial^{a+b} H(q,p)}{\partial p^a\partial q^b}C^{a,b}.
\end{eqnarray}
The equations of motion for the classical moments and expectation values $(q,p)$
are then obtained by computing their Poisson brackets with this Hamiltonian.
The infinite system of equations of motion that is obtained by this procedure
is then completely equivalent to the evolution given by the Liouville equation.

The evolution equations obeyed by the classical moments are
the same as the ones fulfilled by their quantum counterparts with the particularization
$\hbar=0$. These $\hbar$ factors only appear when computing the Poisson brackets between
two moments due to the noncommutativity of the basic operators $\hat q$ and $\hat p$.

In this formalism it is very clear that the classical limit, understood as $\hbar\rightarrow 0$,
of a quantum theory is not a unique trajectory on the phase space, but an ensemble of classical
trajectories described by a probability distribution $\rho$ or its corresponding moments $C^{a,b}$.
In this way, the quantum effects have two different origins. On the one hand, \emph{distributional}
effects are due to the fact that moments can not be vanishing (due to the Heisenberg uncertainty
relation) and generically the centroid of a distribution $(q,p)$ does not follow a classical \emph{point}
trajectory on phase space. (The classical orbit obtained with an initial Dirac delta distribution, for which
all moments vanish, will be referred as classical point trajectory.)
These distributional effects are also present in a classical setting.
On the other hand, there are \emph{noncommutativity} or \emph{purely quantum} effects, which appear as explicit $\hbar$
factors in the quantum equations of motion. These latter effects are due to the noncommutativity
of the basic operators and have no classical counterpart.

The evolution of the classical and quantum moments differ for a generic Hamiltonian due to
the commented $\hbar$ terms. Nevertheless the harmonic Hamiltonians, defined as those
that are at most quadratic in the basic variables, have very special properties and, in particular,
they generate exactly the same evolution in the classical and quantum frameworks. In this paper
the Hamiltonian of a particle on a potential will be studied and, due to these special properties
of the harmonic Hamiltonians, the analysis will be separated between the harmonic and the anharmonic
case. All possible harmonic systems will be studied but, regarding the anharmonic sector,
which is much more involved, only a particular example will be worked out: the pure quartic oscillator.

Once the equations of motion are obtained, the only information left to obtain a dynamical state
are the initial conditions. Nonetheless, the stationary states play a fundamental role in quantum
mechanics. In this setting, moments corresponding to a stationary state can be obtained as
fixed points of the dynamical system under consideration; that is, by dropping all time
derivatives on the equations of motion for $(q,p, G^{a,b})$ and solving the remaining algebraic system.
This system of algebraic equations, as will be made explicit below, is sometimes incomplete
and thus it is not possible to fix the values of all variables $(q,p, G^{a,b})$ of a stationary state
by this method. Nonetheless, as shown in \cite{Ban77, RiBl79, BrSu81}, another condition for the stationary states can be derived as a recursive relation between moments of the form $G^{0,n}$,
by making use of the fact that these states are eigenstates of the Hamiltonian operator
($\langle\hat H\rangle=E$).
For the kind of Hamiltonians that will be treated in this paper, corresponding to mechanical
systems of a particle on a potential $\hat H= \hat p^2/2+V(\hat q)$ with potentials of
the form $V(\hat q)=q^m$ and vanishing expectation value $q$ in its stationary state,
this recursive relation can be written in the following way (see \cite{Bri14} for more details):
\begin{eqnarray}\label{recursive}
(2 k+m+2) G^{0,k+m}&=&2 E (k+1) G^{0,k}\nonumber\\&+&\!\!\frac{\hbar^2}{4}(k+1)k(k-1)G^{0,k-2}.\,\,\,\,\,
\end{eqnarray}
In consequence, whenever moments up to order $G^{0,m}$ are known,
the higher-order fluctuations of the position can be obtained directly.
Classical stationary moments obey this very same equation dropping the last term.

In order to finalize the summary of previous works, let us comment that the moments
corresponding to a valid probability distribution (wave function) are not free
and obey several inequalities. The most simple examples are the non-negativity of
moments with two even indices,
\begin{eqnarray}\label{eveneven}
G^{2n, 2m}\geq 0 \,\,, {\rm for}\,\, n, m\in{\mathbb N},
\end{eqnarray}
and the Heisenberg uncertainty principle,
\begin{equation}\label{heisenberg}
(G^{1,1})^2\leq  G^{2,0}G^{0,2} -\frac{\hbar^2}{4}\,.
\end{equation}
As always, inequalities for classical moments are obtained from the ones of the quantum moments
by taking $\hbar=0$. In Ref. \cite{Bri14} several inequalities for high-order moments
were obtained. These inequalities will be used below to constrain the values of
certain moments of stationary states as well as to monitor the validity of
the numerical resolution of dynamical states.

\section{Particle on a potential}\label{sec_potential}

For definiteness, in order to check the interpretation and applicability
of the formalism for classical and quantum moments summarized in previous
section, here the Hamiltonian for a particle moving on a potential will
be assumed,
\begin{equation}\label{hamiltonian_potential}
\hat H=\frac{\hat p^2}{2}+V(\hat q).
\end{equation}

Let us define the dynamics for the quantum expectation values and moments.
The effective quantum Hamiltonian is given by
\begin{equation}\label{Hq}
H_Q=\frac{p^2}{2}+V(q) + \frac{1}{2}G^{2,0}
+\sum_{n=2}^\infty \frac{1}{n!}\frac{\md^n V(q)}{\md q^n} G^{0,n}.
\end{equation}

From there, it is straightforward to obtain the equations of motion for
the centroid of the distribution:
\begin{eqnarray}\label{eqq}
\frac{\md q}{\md t} &=& p,\\\label{eqp}
\frac{\md p}{\md t} &=& -V'(q)-\sum_{n=2}^\infty \frac{1}{n!}\frac{\md^{n+1} V(q)}{\md q^{n+1}} G^{0,n}.
\end{eqnarray}

Note that the evolution equation of the position $q$ is not modified by
the moments. On the contrary, the equation of motion for its conjugate momentum
$p$ does receive corrections due to the presence of the moments $G^{0,a}$
in the right-hand side of Eq. (\ref{eqp}). It is straightforward to see that the
Hamiltonian $H_C$, which would describe the evolution of a classical distribution
on the phase space, it is obtained by replacing the quantum moments $G^{a,b}$
by the classical ones $C^{a,b}$ in Eq. (\ref{Hq}). The centroid of that
classical distribution will follow the evolution given by (\ref{eqq}-\ref{eqp}),
replacing again $G^{a,b}$ by $C^{a,b}$.

It is enlightening to combine last two equations
in order to obtain the corrected Newton equation,
\begin{equation}
\frac{\md^2 q}{\md t^2}= -V'(q)-\sum_{n=3}^\infty \frac{1}{n!}\frac{\md^{n} V(q)}{\md q^{n}} G^{0,n-1}.
\end{equation}
The moment terms that appear in this
modified equation are sometimes referred as the quantum
contributions to the Newton equations. Nevertheless, we see from our analysis that
the equations of motion for a centroid of a classical distribution in the phase
space characterized by moments $C^{a,b}$ will obey this very same
equation. Therefore this equation must be understood as the fact that the centroid of
a distribution does not follow a classical point trajectory.

Taking the Poisson brackets between moments $G^{a,b}$ and the Hamiltonian (\ref{Hq}),
and separating the terms with an explicit dependence on $\hbar$,
the equations of motion for the quantum moments $G^{a,b}$ can be written as
\begin{widetext}
\begin{eqnarray}
\frac{\md G^{a,b} }{\md t}&=& b\, G^{a+1,b-1}
\!+\!a \sum_{n=2}^\infty \frac{V^{(n)}(q)}{(n-1)!}
\left[G^{0,n-1}G^{a-1,b}-G^{a-1,b+n-1}\right]
\nonumber \\\label{eqG}
&-&\sum_{n=3}^\infty\sum_{k=1}^{M} \frac{V^{(n)}(q)}{(n-2k-1)!} \left(\!\!\begin{array}{c}
a\\2k+1
\end{array}\!\!\right)\left(-\frac{\hbar^{2}}{4}\right)^k
G^{a-2k-1,b+n-2k-1},\label{momenteq_potential}
\end{eqnarray}
\end{widetext}
with $M$ being the integer part of $[{\rm Min}(a,n)-1]/2$. The evolution equation
for the classical moments can be formally obtained from last equation by replacing
all $G^{a,b}$ by $C^{a,b}$ moments and imposing $\hbar=0$, that is, removing all terms
that appear in the second line:
\begin{eqnarray}\label{eqC}
\frac{\md C^{a,b} }{\md t}&=& b\, C^{a+1,b-1}
 \\ \nonumber
&+&a \sum_{n=2}^\infty \frac{V^{(n)}(q)}{(n-1)!}
[C^{0,n-1}C^{a-1,b}-C^{a-1,b+n-1}].
\end{eqnarray}

In summary, Eqs. (\ref{eqq}) and (\ref{eqp}), in combination with (\ref{eqG}), form an infinite closed
system of ordinary differential equations that describes the quantum dynamics
of a particle on a potential $V(q)$ and are completely equivalent to the
Schr\"odinger flow of quantum states (or the Heisenberg flow of quantum operators).
On the other hand, the infinite system composed by Eqs.
(\ref{eqq}), (\ref{eqp}) [replacing $G^{0,n}$ terms by $C^{0,n}$], and (\ref{eqC}) describes
the classical evolution of a probability distribution on the phase space, which
is equivalent to the Liouville equation.

As can be seen in these equations of motion, for a generic potential $V(q)$, all orders couple.
Hence, in order to make these equations useful for a practical purpose, it
is necessary to introduce a cutoff by hand, and assume $G^{a,b}$ to be vanishing
for all  $a+b>N_{\rm max}$, $N_{\rm max}$ being the maximum order to be considered. In order to impose
this cutoff, due to the special properties of the Poisson brackets between two
moments, care is needed (see \cite{Bri14} for a more detailed discussion). In order to truncate properly
the system at an order $N_{\rm max}$, taking into account all contributions up to this order, it is
straightforward to see that the upper limit of the summation in Eq. (\ref{eqp}) should be taken as $N_{\rm max}$.
Regarding the equation for the moments (\ref{eqG}), the sum of the first line
should clearly go up to $(N_{\rm max}+1)$ for the quadratic term in moments, but only up to $(N_{\rm max}-a-b-2)$ for the
second linear term. The summations in the second line of that equation are more involved
and should be replaced by
\begin{equation}
\sum_{n=3}^\infty\sum_{k=1}^M\longrightarrow\sum_{n=3}^{n_{\rm max}}\sum_{k=k_{\rm min}}^M,
\end{equation}
with $n_{\rm max}=N_{\rm max}+a-b$ and, for every fixed $n$, $k_{\rm min}$ the maximum between 1 and
$\lceil (a+ b + n - N_{\rm max} - 2)/4)\rceil$. For the classical equations, the same limits
as in their corresponding quantum equations should be imposed.

The validity of this cutoff should
be proved {\it a posteriori} by solving the equations of motion with
different cutoffs and checking that the solution converges with the cutoff order.

If an integer $N_{\rm max}$ exists, for which $V^{(n)}(q)$ vanishes
for all values $n>N_{\rm max}$, the infinite sums on the right-hand side 
of Eq. (\ref{eqp}) will become finite. Regarding the quantum $G^{a,b}$ (\ref{eqG}) and the classical moments $C^{a,b}$ (\ref{eqC})
of order $a+b$, the highest order that appears in their corresponding
equations of motion is of order $(a+b+N_{\rm max}-2)$.
Therefore, only in the case that $N_{\rm max}\leq 2$ the introduction
of a cutoff will not be necessary. This is in fact the case of
a harmonic Hamiltonian, which will
be analyzed in the next section.

\section{Harmonic potentials: $V'''(q)=0$}\label{sec_harmonic}

The harmonic Hamiltonians $H(q,p)$ are defined as those for which all derivatives
with respect to the basic variables $(q,p)$ higher than second order vanish.
In the case of a Hamiltonian of a particle
on a potential (\ref{hamiltonian_potential}), this happens when $V''(q)=:\omega^2$ is a constant.

This kind of Hamiltonians has very special properties, which were analyzed in Ref. \cite{Bri14}.
Let us briefly summarize its main properties.
First, for this kind of Hamiltonians, equations at every order decouple
from the rest of the orders. Second, equations of motion of expectation values $(q,p)$ do not get any
correction from moment terms and thus there is no back-reaction. Hence,
the centroid of the distribution follows a classical point trajectory.
In addition,
given the same initial data, classical and quantum moments have exactly the same
evolution since no $\hbar$ term appears in the equations of motion. As will be
shown in this section, classical and quantum stationary states differ because
the equations of motion do not provide the complete information to fix the
value of all moments and thus recursive relation (\ref{recursive}) will have to be used.

Due to the mentioned properties, most of the analysis of this section applies equally to classical as well
as to quantum moments. Thus the whole analysis will be performed for quantum moments and
emphasis will be made in the particular points where the situation is different
for classical moments.

The expectation value of a Hamiltonian of a particle on a potential $V(q)$, such that
$V''(q)=\omega^2$ is a constant value, can be written in the following way in terms of
expectation values and moments:
\begin{equation}
H_Q=\frac{p^2}{2}+\frac{\omega^2}{2} q^2+\frac{1}{2}G^{2,0}+\frac{\omega^2}{2}G^{0,2}.
\end{equation}
This is, as explained in previous section, the effective quantum Hamiltonian that
can be used to obtain the equations of motion. In particular, the equations of
motion for the expectation values $q$ and $p$ reduce to their usual form,
\begin{eqnarray}\label{dqdt_harmonic}
 \frac{{\rm d}q}{{\rm d} t}&=&p,\\\label{dpdt_harmonic}
 \frac{{\rm d}p}{{\rm d} t}&=&-V'(q).
\end{eqnarray}
Here it can be seen that, as already commented above,
there is no back-reaction of moments in the equations for
the centroid, in such a way that the centroid follows
a classical phase space orbit.

The equations for the moments (\ref{momenteq_potential}) reduce to,
\begin{equation}\label{dGdt_harmonic}
\frac{\md G^{a,b} }{\md t}= b\, G^{a+1,b-1}
-a \omega^2 G^{a-1,b+1}.
\end{equation}
The classical moments $C^{a,b}$ fulfill this very same equation, replacing
all quantum moments $G^{a,b}$ by their classical counterparts $C^{a,b}$, as can
be readily checked from (\ref{eqC}).

As it is well known, it is not necessary to solve Eqs. (\ref{dqdt_harmonic}--\ref{dpdt_harmonic})
explicitly to obtain the phase-space orbit that is followed by the centroid.
It is sufficient to divide both equations to remove the dependence on time and integrate the resulting equation.
This procedure leads to the implicit solution,
\begin{equation}\label{centroid_orbit}
E_{\rm centroid}=p^2/2+V(q),
\end{equation}
$E_{\rm centroid}$ being the integration constant that parametrizes different orbits,
which can obviously be interpreted as the energy of the centroid.
Note that this $E_{\rm centroid}$ energy is not the expectation value of the Hamiltonian $H_Q$.
In particular, since $H_Q$ (and for the classical treatment $H_C$)
is also a constant of motion, the difference between both, leads to another conserved quantity
in terms of second-order moments: $G^{2,0}+\omega^2 G^{0,2}$ (and $C^{2,0}+\omega^2 C^{0,2}$ for the classical moments).

The first derivative of the potential $V'(q)$ only appears
in the evolution equation for the momentum $p$ (\ref{dpdt_harmonic}) and, certainly, the phase-space
orbit followed by the centroid (\ref{centroid_orbit}) depends on the precise form of the potential.
Nonetheless, note that the equations of the moments (\ref{dGdt_harmonic}) only depend on the second derivative of
the potential $\omega^2$. Therefore, in order to fully analyze the evolution
of the moments, the study will be split
in the two possible and physically different cases: $\omega^2=0$ and
$\omega^2\neq0$. The former describes a particle
moving under a uniform force, whereas the latter corresponds to the harmonic
($\omega^2>0$) and the inverse harmonic ($\omega^2<0$) oscillators.

\subsection{Particle under a uniform force: $V''(q)=\omega^2=0$}

In this subsection the generic linear potential $V=\beta q+V_0$ will be analyzed.
Without loss of generality, $V_0$ will be chosen to be vanishing.
This potential represents a particle under a constant force. The case of a free
particle $(\beta=0)$ will also be included in the analysis.

As explained above, in this case all orders decouple and the centroid of the distribution
follows a classical point trajectory in phase space: $\beta q+p^2/2=E_{\rm centroid}$, with
$E_{\rm centroid}$ a constant value.
Since the full Hamiltonian $H_Q$ is also a constant of motion, it is obvious then that the moment
$G^{2,0}$ is also constant during the evolution. In fact, looking at the equations of motion (\ref{dGdt_harmonic}),
it is immediate to see that the fluctuations of the momentum at all orders $G^{a,0}$
are constants of motion.

Let us first analyze the stationary states, that is, the fixed points of the dynamical system.
Dropping all time derivatives in the system of equations (\ref{dqdt_harmonic}--\ref{dGdt_harmonic}),
it is straightforward to see that only the free
particle ($\beta=0$) case allows for stationary solutions that would be given
by $p=0$ (particle at rest) and all moments $G^{a,b}$ vanishing for all $a\geq1$
and $b\geq 0$. The position $q$ and its fluctuations at all orders $G^{0,b}$
could, in principle, take any value. That is, the particle can be anywhere and with an
unbounded uncertainty in its position. Nonetheless, even if this choice of moments
is valid for the classical case, it is not for the quantum case since it violates
the Heisenberg uncertainty relation (\ref{heisenberg}).
Therefore as it is well known, and contrary
to the classical case, no stationary state can be constructed for the
free quantum particle.

The analytical solution for a dynamical state can be found explicitly
for the evolution of all moments,
\begin{equation}\label{linearsol}
G^{a,b}(t)=\sum_{n=0}^b
\left(
\begin{array}{c}
b\\n
\end{array}
\right)
(t-t_0)^{b-n}G^{a+b-n,n}_0,
\end{equation}
for initial data $G^{a,b}_0:=G^{a,b}(t_0)$. The evolution of the moments
is independent of the value of $\beta$, thus this solution
is valid both for the case of the free particle and the particle under
a uniform force.
As can be seen, each moment is given by a linear combination of the initial
value of the moments of its corresponding order with polynomial coefficients
on the time parameter. The state spreads away from its initial configuration
and, for large times, the moments $G^{a,b}$ increase as $t^b$.
The initial conditions of this state are still free. For instance, it is possible to
choose an initial state of minimum uncertainty but, even so, all moments,
except the constants of motion $G^{a,0}$, will increase with time.

\subsection{Harmonic and inverse harmonic oscillators: $V''(q)=\omega^2\neq 0$}

It is well known that any potential of the form $V=\frac{\omega}{2} \tilde q^2+\beta \tilde q+V_0$
can be taken to the form $V=\frac{\omega^2}{2}q^2$ by a shift of the
variable $q=\tilde q+\frac{\beta}{\omega^2}$ and a redefinition of
the value of the potential at its minimum ($V_0=\frac{\beta^2}{2\omega^2}$),
which does not have any physical meaning. If $\omega^2$ is positive, this is the potential
of a harmonic oscillator, a ubiquitous system in all branches of physics. Since
the equations of motion for expectation values (\ref{dqdt_harmonic}--\ref{dpdt_harmonic}) do not get any backreaction by
moments, their solutions are oscillatory functions and they follow an elliptical orbit
in phase space. On the other hand, the case $\omega^2<0$ corresponds to the
inverted harmonic oscillator. This system can be viewed as an oscillator with
imaginary frequency. The solution for the expectation values $(q,p)$ are hyperbolic functions
and they follow hyperbolas in phase space. In the rest of this subsection the behavior
of the moments will be considered for both systems.

Let us first analyze the stationary states. Equaling to zero the right-hand side of the equations of motion
(\ref{dqdt_harmonic}--\ref{dGdt_harmonic}),
the equilibrium point $p=0=q$ for the expectation values, as well as the recursive
relation $b\,G^{a+1,b-1}=a\omega^2 G^{a-1,b+1}$ for the moments are obtained.
The solution to this recursive relation is given by the following condition
for moments with both indices even numbers,
\begin{equation}\label{recsol1}
G^{2a,2b}=
\frac{2a!\,2b!}{(2(a+b))!}\frac{\left(a+b\right)!}{a!\,b!}
\omega^{2a} G^{0,2(a+b)},
\end{equation}
whereas the rest of the moments must vanish.
If the sign of $\omega^2$ was negative, that would impose
some moments with even indices to be negative. This is not
acceptable since all moments of the form $G^{2a,2b}$ are
non-negative by construction (\ref{eveneven}). Thus, from here it is immediately concluded
that the inverse oscillator can not have stationary states.

As can be appreciated in the last relation (\ref{recsol1}), even if the information
concerning the stationary state contained in the equations of motion has been exhausted,
there is still one freedom left at each order. This freedom is represented
in this equation by the high-order fluctuations of the position $G^{0,n}$.

In order to fix the moments $G^{0,n}$, the recursive relation
(\ref{recursive}) can be made use of. For the potential under consideration, that relation reads
\begin{equation}\label{oscillatorrec}
\omega^2 \,(k+2)\, G^{0,k+2}=2(k+1) E G^{0,k}+\frac{\hbar^2}{4}(k+1)k(k-1) G^{0,k-2}.
\end{equation}
This last equation allows us to compute all $G^{0,n}$ moments as function of the energy
at the stationary point $E=(G^{2,0}+\omega^2 G^{0,2})/2=G^{2,0}$
and Planck constant $\hbar$. Taking the limit $\hbar\rightarrow 0$, the (two point) recursive
relation obeyed by classical moments is obtained, which can be easily solved. Combining this solution with
(\ref{recsol1}), the classical moments corresponding to a stationary situation
of the harmonic oscillator can be written in a closed form. Those with two even indices
read
\begin{equation}\label{oscillatorcmoments}
C^{2a,2b}=\frac{(2a)!(2b)!}{a!b!(a+b)!}\frac{E^{a+b}}{2^{a+b}\omega^{2b}},
\end{equation}
and the rest are vanishing.

The quantum case is a little bit more involved. The second-order moments
$G^{2,0}$ and $G^{0,2}$ have the same form as their classical counterparts
in terms of the energy $E$ and the frequency $\omega$ (\ref{oscillatorcmoments}).
But higher-order moments will take corrections as a power series in the
parameter $\hbar^2$ when solving the recursive relation (\ref{oscillatorrec}).
Here we give the explicit expression of all the fluctuations of the position $G^{0,n}$
up to order ten:
\begin{eqnarray*}
G^{0,2}&=& \frac{E}{\omega^2},\\
G^{0,4}&=& \frac{3 }{2}\left(\!\frac{E}{\omega^2}\!\right)^2\!+\frac{3 }{8}\left(\!\frac{\hbar}{ \omega}\!\right)^2 ,\\
G^{0,6}&=& \frac{5}{2}\left(\!\frac{ E}{ \omega ^2}\!\right)^3\!+\frac{25}{8}\left(\!\frac{E}{ \omega^2}\!\right)\!\left(\!\frac{ \hbar}{ \omega}\!\right)^2,\\
G^{0,8}&=& \frac{35}{8}\left(\!\frac{ E}{ \omega ^2}\!\right)^4\!\!+
\!\frac{245}{16}\left(\!\frac{ E}{\omega ^2}\!\right)^2\left(\!\frac{\hbar}{\omega}\!\right)^2\!\!
+\frac{315}{128}\left(\!\frac{ \hbar}{ \omega}\!\right)^4,\\
G^{0,10}&=& \frac{63}{8}\left(\!\frac{ E}{ \omega ^{2}}\!\right)^5\!\!
+\frac{945}{16}\left(\!\frac{ E }{\omega ^2}\!\right)^3\!\!\left(\!\frac{\hbar}{\omega}\!\right)^2\!\!
+\frac{5607}{128}\left(\!\frac{ E}{ \omega ^2}\!\right)\!\!\left(\!\frac{ \hbar}{ \omega}\!\right)^4\!\!\!.
\end{eqnarray*}
The rest of the nonvanishing moments are proportional to these and can be obtained by using the solution (\ref{recsol1}).
Note that a quantum moment $G^{a,b}$ is equal to its classical counterpart (\ref{oscillatorcmoments})
plus certain corrections that are given as an even power series in $\hbar$. This power series goes from $\hbar^2$
up to $\hbar^{2n}$, $n$ being the integer part of $(a+b)/4$.

The only information that is left here is the exact
form of the energy spectrum: $E=\hbar\omega (n+1/2)$. This
is the only input needed in order to obtain the complete realization of the system.
In fact, one could obtain all the moments corresponding to the ground state
by assuming that it is an unsqueezed state with minimum uncertainty that
saturates the Heisenberg relation (\ref{heisenberg}), $G^{2,0}G^{0,2}=\hbar^2/4$, which
implies $E_{\rm ground}=\hbar\omega/2$. In addition note that, as expected,
for this ground state the expression of the quantum moments reduces to the moments
corresponding to a Gaussian probability distribution with width $\sqrt{\hbar/\omega}$.
[The explicit expression for the moments of a Gaussian state is given below (\ref{gaussianG}).]

Regarding the dynamical states, it is easy to solve the equations of motion (\ref{dqdt_harmonic}--\ref{dGdt_harmonic}).
The solution for the moments $G^{a,b}$ can be written as a linear combination of functions
of the form $e^{\pm i \alpha\omega t}$. For moments of even orders, $a+b=2n$, $\alpha$ takes even
values: $\alpha= 0,2,\dots, 2n$; whereas for those of odd orders, $a+b=2n+1$, it takes odd
values: $\alpha= 1,3,\dots, 2n+1$. Thus, the dynamical behavior of the harmonic oscillator ($\omega^2>0$)
and the inverse oscillator ($\omega^2<0$) is completely different.
For the oscillatory case ($\omega^2>0$), all moments $G^{a,b}$ are bounded and they are oscillating
functions. On the contrary, the moments corresponding to the inverse oscillator
are exponentially growing and decreasing functions of time.

\section{The anharmonic case: the pure quartic oscillator}\label{sec_anharmonic}

The potential of the pure quartic oscillator is given by
\begin{equation}
V(q)=\lambda \,q^4,
\end{equation}
which leads to an effective Hamiltonian of the form
\begin{equation}\label{hamiltonian_quartic}
 H_Q= \frac{p^2}{2} + q^4 \lambda+ \frac{1}{2} G^{2, 0} + 6 q^2 \lambda G^{0, 2} + 
 4 q \lambda G^{0, 3} + \lambda G^{0, 4}.
\end{equation}
From this Hamiltonian it is easy to get the equations of motion for the expectation
values,
\begin{eqnarray}\label{qquartic}
 \frac{\md q}{\md t}&=& p,\\\label{pquartic}
 \frac{\md p}{\md t}&=& -4\lambda (q^3+3 q G^{0,2} + G^{0,3}),
 \end{eqnarray}
and for the moments
\begin{eqnarray}\label{eq_quartic}
\frac{\md G^{a,b}}{\md t}&=&b \,G^{a+1,b-1}
+4 \,a\, \lambda\,  [3 \, q \,G^{0,2}+G^{0,3}] G^{a-1,b}
\nonumber\\&-&4 \,a \,\lambda  \left[3\,  q^2\, G^{a-1,b+1}+3\, q \, G^{a-1,b+2}+G^{a-1,b+3}\right]
\nonumber\\
&+&a \,\hbar^2 \,\lambda \,(a-2)\, (a-1)\,   [q\,G^{a-3,b}+G^{a-3,b+1}].
\end{eqnarray}
As can be seen, in this case all orders couple. More specifically, in the equation for a moment
$G^{a,b}$ there appear moments of order two, three and of all orders from ${\cal O}(a+b-3)$ to ${\cal O}(a+b+2)$.

The centroid of a classical distribution will follow the same equations (\ref{qquartic}) and
(\ref{pquartic}), replacing moments $G^{a,b}$ by their classical counterparts,
\begin{eqnarray}\label{qquarticc}
 \frac{\md q}{\md t}&=&p ,\\\label{pquarticc}
 \frac{\md p}{\md t}&=& -4\lambda (q^3+3 q C^{0,2} + C^{0,3}),
 \end{eqnarray}
whereas the evolution of the classical moments will be given by
\begin{eqnarray}\label{eq_quarticc}
\frac{\md C^{a,b}}{\md t}&=&b \,C^{a+1,b-1}
+4 \,a\, \lambda\,  [3 \, q \,C^{0,2}+C^{0,3}] C^{a-1,b}
\\\nonumber&-&4 \,a \,\lambda  \left[3\,  q^2\, C^{a-1,b+1}+3\, q \, C^{a-1,b+2}+C^{a-1,b+3}\right].
\end{eqnarray}
The explicit order coupling differs a little bit from the quantum case, since in this
equation there are only moments of order two, three and of all orders between ${\cal O}(a+b-1)$ and ${\cal O}(a+b+2)$.
 
\subsection{Stationary states}

In order to obtain the stationary states of the pure quartic oscillator,
the infinite set of algebraic equations obtained by equaling to zero the right-hand
side of Eqs. (\ref{qquartic}--\ref{eq_quartic}) must be solved.
Furthermore, recursive relation (\ref{recursive}) must also be obeyed. In this particular case, that relation takes the following form:
\begin{equation}\label{relat}
2\lambda (a + 3) G^{0, a + 4} = 2 E (a + 1) G^{0, a} + \frac{\hbar^2}{4} (a + 1) a (a - 1) G^{0, a - 2},
\end{equation}
with the energy given by the numerical value of the expectation value of the Hamiltonian,
\begin{equation}\label{energy_quartic}
E=H_Q. 
\end{equation}

In practice, due to the coupling of the system, it is necessary to introduce a cutoff
in order to get a finite system and be able to solve it. In our case different
cutoffs have been considered (specifically $N_{\rm max}$= 15, 20, 25, and 30) and
the mentioned system of equations, in combination with relation (\ref{relat}) and
the definition of the energy (\ref{energy_quartic}), has been analytically solved.
The idea behind performing this computation for several cutoffs is to
study the convergence of the solution, that is, to check whether the solution
for the moments does not change when considering higher-order cutoffs.

In principle, there are two different solutions: one that corresponds to the
classical stationary configuration (and thus its equilibrium position is at
the origin $q=0$) and another, for which
the position must not be vanishing [note that this is possible due to the moment
terms that appear in the Hamilton equation (\ref{pquartic}) ] and does not have a classical
point counterpart. Nevertheless, for this latter case, the solution for some moments
with both even indices turns out to be negative, which makes this solution invalid.
Therefore, and as one would expect from symmetry considerations, the expectation values
of any stationary state of the quartic oscillator corresponds to the origin of the
phase space ($p=0=q$). Furthermore, it can be seen that all its corresponding moments
$G^{a,b}$ are vanishing in case any of the indices $a$ or $b$ is an odd number.
The remaining moments can be written in terms of the energy $E$ and the fluctuation
of the position $G^{0,2}$, or any other chosen moment. That is, there is not enough
information in our system of equations to fix all moments and one of them is free.

Regarding the convergence of the solution, comparing the solution obtained with the cutoff $N_{\rm max}=30$
with the one corresponding to $N _{\rm max}=15$, we see that
the expression of all moments coincides up to order 8, whereas the solution with
$N_{\rm max}=30$ and $N_{\rm max}=20$ give the same expression for all moments
up to order 12. Finally, solutions that correspond to $N_{\rm max}=30$ and $N_{\rm max}=25$
coincide up to order 14. From here the existence of a clear convergence of the solution with the
cutoff order is concluded. Nevertheless, this convergence seems to be slower with higher orders.
Here the explicit expressions for all nonvanishing moments up to sixth order is provided:
\begin{eqnarray}
\nonumber
G^{2,0}&=& \frac{4}{3}E,\\\nonumber
G^{4,0}&=& \frac{2}{7} \left(8 E^2+15\hbar^2 \lambda  G^{0,2}\right),\\\nonumber
G^{2,2}&=&\frac{1}{5} \left(4 E G^{0,2}+\hbar^2\right),\\\nonumber
G^{0,4}&=&\frac{1}{3 \lambda }E,\\\nonumber
G^{6,0}&=& \frac{10}{77} \left(32 E^3+228 E \hbar^2 \lambda  G^{0,2}+21\hbar^4 \lambda \right),
\\\nonumber
G^{4,2}&=& \frac{2}{45} E\left(24 E G^{0,2}+41 \hbar^2\right),
\\\nonumber
G^{2,4}&=&\frac{4}{21 \lambda} E^2+\frac{6}{7} \hbar^2 G^{0,2},
\\ G^{0,6}&=& \frac{3}{20 \lambda } \left(4 E
   G^{0,2}+\hbar^2\right).\label{momentsquartic}
\end{eqnarray}
The classical moments $C^{a,b}$, as always, take the same values as their quantum counterparts
with the particularization $\hbar=0$. In these expressions the singular behavior of the limit
$\lambda\rightarrow 0$ is made explicit as the divergence of several moments. This fact does not allow
to perform regular perturbative treatments of this system.

In summary, after imposing the stationarity condition on Eqs. (\ref{qquartic}--\ref{eq_quartic})
and using the definition of the energy (\ref{energy_quartic}) in combination with the recursive relation (\ref{relat}),
the only information left in order to characterize completely any stationary
state of the pure quartic oscillator is the energy $E$ and the fluctuation of
the position $G^{0,2}$.

In addition to these equations already mentioned,
there is still some information more than we can get by making use of the inequalities
obtained in Ref. \cite{Bri14}. In the following, use will be made of those
relations to constrain the values of $G^{0,2}$ and the energy $E$.
For instance, Heisenberg uncertainty principle (\ref{heisenberg}) provides
a lower bound for the product between $E$ and $G^{0,2}$:
\begin{equation}
\frac{3 \hbar^2}{16}\leq E G^{0, 2}.
\end{equation}
Higher-order inequalities give more complicated relations,
which must be fulfilled by the energy $E$
and the fluctuation of the position $G^{0,2}$ of any stationary state
of this system.

For the particular case of the ground state a reasonable assumption is that,
as happens for the harmonic oscillator,
it saturates the above relation. This would give
$G^{0,2}_{\rm ground}=3\hbar^2/(16 E_{\rm ground})$ and let the energy
of the ground state $E_{\rm ground}$ as the only unknown physical quantity in (\ref{momentsquartic}).
Introducing then these expressions of the moments of the ground state
in terms of $E_{\rm ground}$ in the higher-order inequalities, an upper and lower bound for the energy is obtained.
By considering inequalities that only contain moments up to fourth-order
yields the following result:
\begin{equation}
\frac{3}{4}\left(\frac{45}{68}\right)^{1/3}
 \leq  \frac{E_{\rm ground}}{(\hbar^4 \lambda)^{1/3}} \leq
 \frac{1}{4}\left(\frac{85}{4}\right)^{1/3},
 \end{equation}
or, in decimal notation,
\begin{equation}\label{interval4}
0.654
\leq \frac{E_{\rm ground}}{(\hbar^4 \lambda)^{1/3}} \leq
0.692,
\end{equation}
which already provides a good constraint on the energy. Furthermore,
all inequalities that contain moments up to order six reduce to
the following tighter interval of validity for the energy:
\begin{equation}
\frac{3}{4}\left(\frac{45}{68}\right)^{1/3}
 \leq  \frac{E_{\rm ground}}{(\hbar^4 \lambda)^{1/3}} \leq
 \frac{9}{4}\left(\frac{3}{116}\right)^{1/3},
 \end{equation}
or, writing these fractions as decimal numbers,
\begin{equation}\label{interval6}
0.654
\leq \frac{E_{\rm ground}}{(\hbar^4 \lambda)^{1/3}} \leq
0.665.
\end{equation}
This gives a very tight constraint on the energy of this bound state.
Nevertheless, the exact (numerically computed) energy of this state is available
in the literature (see e.g. \cite{BOW77, Vor94}): $E_{\rm ground}=0.670039(\hbar^4 \lambda)^{1/3}$
\footnote{Note that in the mentioned references the considered kinetic term in
the Hamiltonian is chosen to be $p^2$, instead of $p^2/2$. Therefore a
factor $\sqrt[3]{4}$ must be introduced to relate the energy eigenvalues given in those
references with the one obtained by the notation of the present paper.}.
This numerical value is very close but outside the derived interval.
Therefore,
we can conclude that, even if the saturation of the Heisenberg uncertainty
is a reasonable assumption for the ground state that provides a good
estimation of the ground energy, this assumption is not satisfied and
the uncertainty relation is not completely saturated for the present model.

This analysis shows the practical relevance of the inequalities that were derived
in Ref. \cite{Bri14} as a complementary method to extract
physical information from the system. Certainly the inequalities will not
give exact relations between different quantities, but intervals of
validity can be extracted from them.
Finally, let us stress the importance of considering higher-order inequalities.
Note that the interval derived from fourth-order inequalities (\ref{interval4}) does indeed
allow the exact (numerical) value of the ground energy, and thus in principle permits the 
saturation of the uncertainty relation. Therefore,
in this particular example inequalities up to fourth order allowed a
property of the system, which is forbidden by the stronger condition
derived from higher-order ones.

\subsection{Dynamical states}

The classical point trajectory of the pure quartic oscillator, that is, the solution to
Eqs. (\ref{qquartic}-\ref{pquartic}) neglecting all moments, can only
be written in terms on hypergeometric functions. Nevertheless, the orbits
on the phase space are
easily obtained by the conservation of the classical energy: $E_{class}=\frac{p^2}{2}+\lambda q^4$.
Contrary to the harmonic oscillator, the period depends on the
energy $E_{class}$ of the orbit, and it is not a constant for different orbits.
For latter use, note that the maximum (classical) value of the position and the
momentum can be directly related to the energy as
$q_{\rm max}^4=E_{class}/\lambda$ and $p^2_{\rm max}=2E_{class}$.
In order to compare different solutions, below we will also make use of a (squared) Euclidean distance
on the phase space ($p^2+q^2$). The maximum  distance from the origin of a given orbit is
reached at ($(p^2+q^2)_{\rm max}=2E_{class}+1/(8\lambda)$).

We are interested on analyzing the quantum and classical evolution of a distribution
that, respectively, follows Eqs. (\ref{qquartic}-\ref{eq_quartic}) and (\ref{qquarticc}-\ref{eq_quarticc}). Nonetheless,
due to the complicated form of these evolution equations,
the possibility of getting  an analytical solution seems unlikely.
Hence, in order to analyze the dynamics of the system, it is necessary to resort to numerical methods.
Here a comment about notation is in order. When the meaning is not clear from the context,
we will sometimes denote as $q_q(t)$ the solution of the quantum
distributional system (\ref{qquartic}-\ref{eq_quartic}), $q_c(t)$
the solution of the classical distributional system (\ref{qquarticc}-\ref{eq_quarticc}),
and finally $q_{class}(t)$ the solution corresponding to the classical point trajectory, that is,
the solution to Eqs. (\ref{qquartic}-\ref{pquartic}) dropping
all moments. The very same notation will be used for the different solutions of the momentum $p(t)$.

For a numerical resolution of the system, two choices have to be made.
On the one hand, for practical reasons, a cutoff $N_{\rm max}$ has to be
considered in order to truncate the infinite system. On the other hand,
it will be necessary to choose initial conditions for the state to be analyzed.

Regarding the truncation of the system, the dynamical equations
for different values of the cutoff will be considered. More precisely, both the
quantum system (\ref{qquartic}-\ref{eq_quartic}) and the classical distributional
system (\ref{qquarticc}-\ref{eq_quarticc}) for every order up to tenth order
will be solved. In this way, it will be possible to check
the convergence of the solution with the considered $N_{\rm max}$, as well as
study differences between the classical and quantum moments.

Concerning the initial conditions, since the movement of the system is oscillatory
around the equilibrium point $q=0$,
a vanishing value for the initial expectation of the position $q(0)=0$
will be considered
without loss of generality. For the expectation of the momentum $p$, in order to
check the dependence of the properties of the system with the energy,
we will make evolutions for several values, namely $p(0)=10,10^2,$ and $10^3$.
Note that the initial classical (point) energy ($p(0)^2/2$) will not be conserved
through evolution; instead, the complete Hamiltonian (\ref{hamiltonian_quartic}) will
be constant. Nevertheless,
due to the correspondence principle, the larger the classical energy,
a somehow more classical behavior is expected to be found. This can already be inferred
from the equations themselves: in the case that moments are negligible with
respect to expectation values $q$ and $p$, the centroid will approximately
follow a classical point orbit on phase space.

As for the initial values of the fluctuations and higher-order moments, a
peaked state given by a Gaussian of width $\sqrt{\hbar}$ will be chosen.
Its corresponding moments $G^{a,b}$ are vanishing if any of the
indices $a$ or $b$ are odd. The only nonvanishing moments take the following values \cite{BBH11}:
\begin{equation}\label{gaussianG}
G^{2a,2b}=\hbar^{a+b} \frac{2a!\, 2b!}{2^{2(a + b)}\,a!\,b!}.
\end{equation}
Therefore, initially the fluctuation of the position and of the momentum are $G^{0,2}=G^{2,0}=\hbar^2/2$.
In principle the initial conditions for the classical pair $q$ and $p$ should be chosen large
in comparison with their fluctuations, so that we can be safely say that we are in a semiclassical
region where this method is supposed to provide trustable results. Nonetheless,
in this case the system oscillates around $q=0$ and in the turning points the momentum
vanishes $p=0$. Thus, for this case the condition of $q$ and $p$ being much larger than
their corresponding fluctuations can not be a good measure of semiclassicality. We will
check if, as already mentioned above, the classical (point) energy of the system does play
such a role.

Given this setting, we will be interested in analyzing several aspects of the system.
$i/$ The validity of this method based on the decomposition of the classical and quantum
probability distributions in terms of moments. In particular the convergence of the system with
the truncation order $N_{\rm max}$ as well as other control methods, like the conservation
of the full Hamiltonian, will be analyzed.
$ii/$ The dynamical behavior of the moments.
$iii/$ The deviation, due to quantum effects, from the classical trajectory on the phase space.
$iv/$ The relative relevance of the two different quantum effects that have been discussed
in Sec. \ref{sec_formalism}: the distributional ones and the noncommutativity or purely quantum ones.
$v/$ The validity of the correspondence principle. That is,
do systems with a larger energy have somehow a more classical
behavior than those with lower energy?

Regarding the first two question ($i/$ and $ii/$) all results that will be commented
for the quantum moments apply also to the classical ones. Furthermore,
except for the last issue ($v/$) about the correspondence principle, the qualitative behavior
of the system is the same for all considered values of initial momentum $p(0)$.
Hence, the results regarding the first four points ($i/$ to $iv/$) will be presented
for the particular case of $p(0)=10$ and, finally, the last point ($v/$) will be discussed
by comparing results obtained for different initial values of the classical energy. 
In all numerical simulations $\lambda=1$ and $\hbar=10^{-2}$ have been considered.

$i/$  The natural tendency of the both quantum and classical moments is to increase
with time, since the
dynamical states are deformed through evolution. This formalism
is best suited for peaked states so, when higher-order moments become
important, it is expected not to give trustable results. Numerically
this is seen in the fact that,
after several periods, the system becomes unstable and thus the results
are no longer trustable.

In order to check the validity of our results we have several
indicators at hand: numerical convergence of the solution, conservation
of the constants of motion (in this case the full Hamiltonian), convergence
of the results with the order of the cutoff, and fulfillment of the
inequalities derived in \cite{Bri14}. The numerical convergence
has been checked by the usual method: by computing several solutions with
an increasing precision and confirming that the difference between them
and the most precise one tends to zero. The full Hamiltonian has also
been verified to be conserved during the evolutions presented in this paper.

\begin{figure}
 \includegraphics[width=0.48\textwidth]{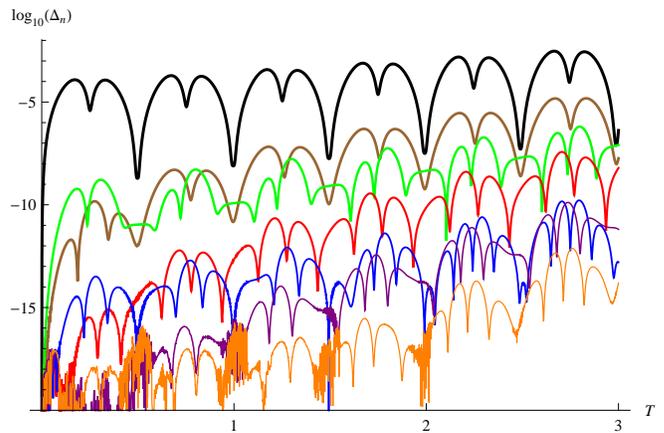}
 \caption{The squared Euclidean distance on phase space between orbits corresponding to consecutive
 orders $\Delta_n:=[q_n(t) - q_{n-1}(t)]^2+[p_n(t) - p_{n-1}(t)]^2$ is shown in a logarithmic plot for $n=2,\dots, 8$.
The distance between the second-order and the classical point trajectory ($\Delta_2$) corresponds to the black (thickest)
line. For the distance corresponding to higher orders ($\Delta_n$), the following colors have been used:
brown ($n=3$), green ($n=4$), red ($n=5$), blue ($n=6$), purple ($n=7$), and orange ($n=8$); the thickness of the lines
being decreasing with the order. The estimated numerical error of these solutions is around $10^{-16}$, thus higher orders
are almost numerical error during the first two periods. Note that, at any time, we get a very rapid and strong convergence
with the considered order.}\label{figconvergence}
\end{figure}

For the analysis of the convergence of the system with the truncation
order, we define the squared Euclidean distance between points on
the phase space as $\Delta_n(t):=[q_n(t) - q_{n-1}(t)]^2+[p_n(t) - p_{n-1}(t)]^2$,
with $q_n(t)$ and $p_n(t)$ being the solution of the system truncated at $n$th order.
In particular $q_1(t)=q_{class}(t)$ and $p_1(t)=p_{class}(t)$ correspond to classical point orbits.
This will serve as a measure of the departure of the solution at every order
from the previous order. In Fig. \ref{figconvergence} the distance $\Delta_n$
between consecutive solutions is drawn in a logarithmic scale for $n=2,\dots, 8$
for the first three periods of the evolution. From this plot it is clear that,
at least during a few periods, the convergence is very fast with the truncation order.
We have not included the $\Delta_9$ and $\Delta_{10}$, since their value is
lower than $10^{-16}$, the estimated numerical error for the solutions shown
in this plot, during the first three periods. It is interesting to note that,
whereas the rest of the $\Delta_n$ have a more complicated structure, $\Delta_2$
(shown by the thickest black line in Fig. \ref{figconvergence}) follows a periodic
pattern with a local minimum every quarter of a period. These points of minimum
deviation from the classical orbit correspond to points with maximum momentum
($q=0$) and to turning points ($p=0$).

\begin{figure}
\includegraphics[width=0.48\textwidth]{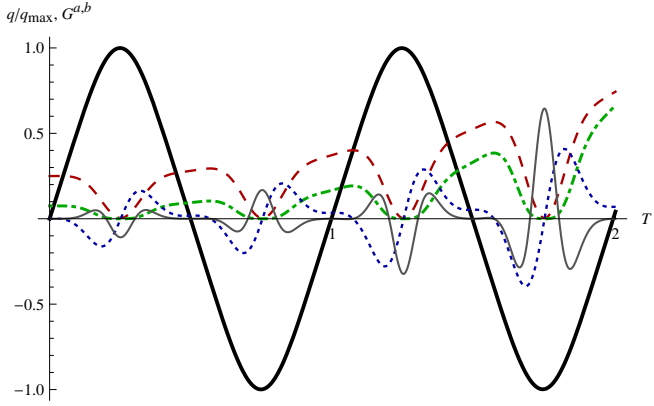}
\caption{In this figure the evolution of the position $q$ over its maximum value $q_{\rm max}$
(black continuous thick line) with respect to to the time
(measured in terms of the period $T$) is shown. The rest of the lines correspond to some moments $G^{a,b}$
rescaled by a factor for illustrational purposes. More precisely, the red (long-dashed) line corresponds to
$50 G^{0, 2}$, the green (dot-dashed) line to $10^3 G^{0, 4}$, the blue (dotted) line to $10 G^{1,1}$,
and the gray (continuous thin) line to $10^2 G^{2,1}$. The behavior of the moments is oscillatory, with an
increasing amplitude.}\label{figqmoments}
\end{figure}

Remarkably we have found that the inequalities are the first indicator to signalize
the wrong behavior of the system. In the particular case with $p(0)=10$, the
tenth-order solution obeys all inequalities that contain only moments
up to fourth order during more than five periods. But some of the inequalities that
contain moments of sixth order are violated soon after the fourth cycle.
Finally, some inequalities with eighth order moments are violated after around 2.5--3
cycles. In fact, it is expected that the values obtained for higher-order
moments be less accurate than those for lower-order ones due to the truncation
of the system. As already commented above, in the evolution equation of
a moment $G^{a,b}$, there appear moments from order ${\cal O}(a+b-3)$
to ${\cal O}(a+b+2)$ [only from ${\cal O}(a+b-1)$ to ${\cal O}(a+b+2)$
for the classical moments]. Thus, when we perform the truncation, let us say, at
order $N_{\rm max}$, we remove several terms from the equations of motion for moments
of order $(N_{\rm max}-1)$ and $(N_{\rm max}-2)$, whereas evolution equations for
lower-order moments are considered in a complete form. Therefore, moments
of order $(N_{\rm max}-1)$ and $(N_{\rm max}-2)$ suffer the presence of the cutoff
directly. On the other hand, lower-order moments only feel the presence of
the cutoff indirectly, due to the coupling of the equations.

In summary, after the analysis explained in the last few paragraphs,
it is quite safe to assert that the results derived during the
first 2.5 cycles are completely trustable [for $p(0)=10$].
As can be seen, in most of the plots only two periods are shown.

$ii/$ The fluctuations and higher-order moments are oscillatory functions
that evolve increasing their amplitude. In Fig. \ref{figqmoments}
the evolution of some moments, as well as of the expectation value
of the position $q$, is shown as an example. Note that, for
illustrational purposes, the moments have been multiplied by
different factors and the position is divided by its (classical)
maximum value $q_{\rm max}$.
Interestingly, moments $G^{0,2}$ and $G^{0,4}$ are almost vanishing
at turning points, when the position takes its maximum value, and
have a maximum soon after $q$ crosses its origin.

\begin{figure}[h]
\begin{center}
\includegraphics[width=0.5\textwidth]{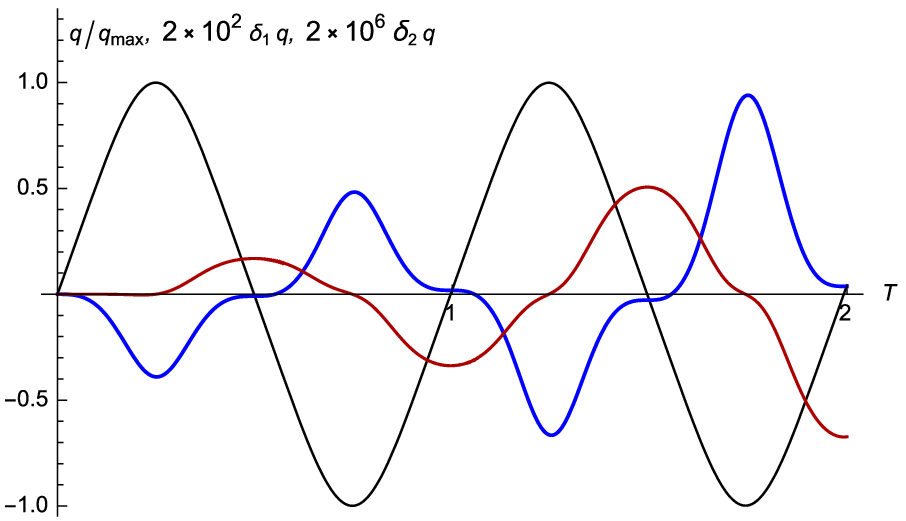}\\\mbox{}
\includegraphics[width=0.5\textwidth]{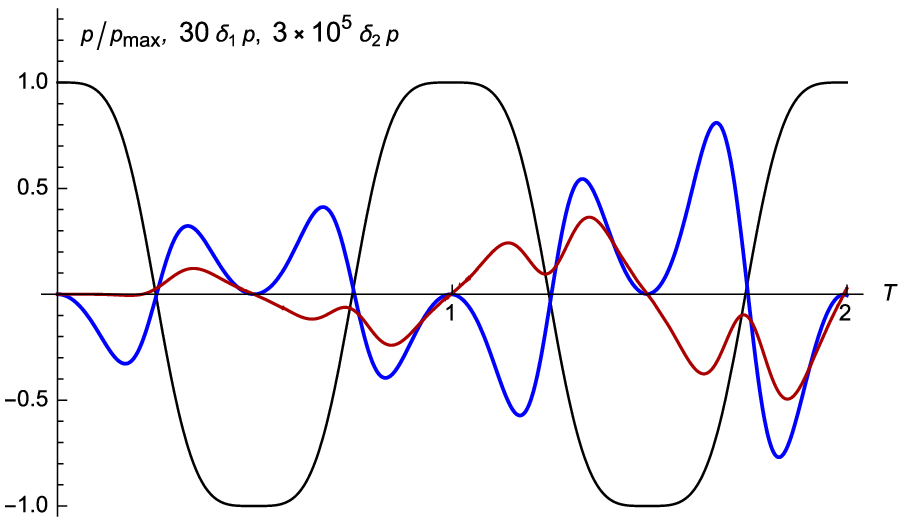}\\\mbox{}
\includegraphics[width=0.5\textwidth]{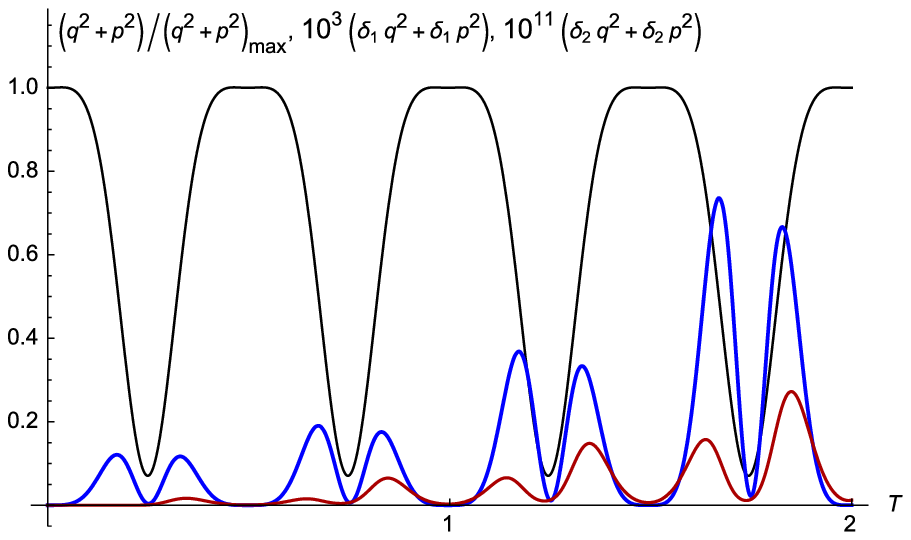}
\end{center}
\caption{In these plots the evolution of $q$, $p$,
and $(q^2+p^2)$ (divided by their maximum values)
is shown in combination with the operators $\delta_1$ and $\delta_2$ acting
on them. The black (thinnest) line represents the evolutions of
the quantity we are considering, for instance in the upper plot
$q/q_{\rm max}$, the blue (thickest) line represents the distributional
effects, in the mentioned plot $\delta_1 q$, whereas the red line stands for
the purely quantum effects, in the considered graphic $\delta_2q$.}\label{figp010}
\end{figure}

$iii/$ and $iv/$ In order to analyze the deviation of the quantum and
classical distributional trajectories from their corresponding classical
point orbit, two operators, $\delta_1$ and $\delta_2$, are defined as follows
\begin{eqnarray}
\delta_1 q(t) &:=&q_{c}(t)-q_{class}(t) ,\\
\delta_2 q(t)&:=&q_{q}(t)-q_{c}(t).
\end{eqnarray}
The same definitions apply for $\delta_1p$ and $\delta_2p$.
These operators are a measure of the two
quantum effects that were defined in \cite{Bri14} and have been discussed
in Sec. \ref{sec_formalism} of the present paper.
On the one hand, the operator $\delta_1$ will contain the strength
of the distributional effects. On the other hand, $\delta_2$
will encode the intensity of purely quantum effects, whose
origin is  due to the $\hbar$ factors that appear explicitly
in the quantum equations of motion.
In our numerical analysis $q_c(t)$ and $q_q(t)$ will be considered to be
the solutions to the corresponding truncated system at order 10.
Finally, the complete departure from the classical orbit will
be given by the sum of both differences:
\begin{equation}
\delta q=\delta_1 q+\delta_2 q=q_{q}-q_{class}.
\end{equation}

Figure \ref{figp010} shows the evolution of the system
as well as the differences given by the operators $\delta_1$ and $\delta_2$
acting on different variables in terms of time.
(Note that these differences are multiplied by certain enhancement factors for
illustrational purposes.) More precisely, in the upper plot of the mentioned figure
the evolution of the position divided by its (classical) maximum $q/q_{\rm max}$,
as well as the differences $\delta_1q$ and
$\delta_2q$, are shown. The middle plot represents the evolution of $p/p_{\rm max}$
with its corresponding $\delta_1p$ and $\delta_2p$. Finally, in the lower graphic
the squared Euclidean distance from the origin of the phase space is plotted, as well as
the deviations $(\delta_1p^2+\delta_1q^2)$ and $(\delta_2p^2+\delta_2q^2)$
\footnote{All objects of the form $\delta x^2$ must be understood as $(\delta x)^2$.}.
This
distance has been divided by its maximum classical value which, as commented above,
can be easily related to the initial conditions as $(p^2+q^2)_{\rm max}=p(0)^2+1/(8\lambda)$.

Looking at the enhancement factors that have been introduced for the differences
$\delta_1$ and $\delta_2$ so that objects that have been plotted appear approximately
with the same order of magnitude, it is straightforward to see that for all quantities the departure
from the classical point trajectory $\delta$ is mainly due to the distributional effects
measured by $\delta_1$.
In particular, during the two cycles that are shown, the absolute maximum departure from the
classical trajectory is of the order of $\delta q\approx\delta_1q\approx5\times10^{-3}$
for the position and $\delta p\approx\delta_1p\approx 3\times 10^{-2}$
for the momentum. Combining this result, it is direct to obtain the maximum departure as 
measured by the squared Euclidean distance on the phase space: $\delta q^2 +\delta p^2\approx 10^{-3}$;
which also can be obtained from the lower plot of Fig. \ref{figp010}.

As already commented, and as one of the main results of this paper, in this model
we have shown that the distributional effects are much more relevant than
the purely quantum ones. Let us analyze its relative importance: from the values
that can be seen in Fig. \ref{figp010} we have that $\delta_2q/\delta_1q\approx\delta_2p/\delta_1p\approx10^{-4}$.
This ratio happens to be of the order of $\hbar^2$, which is a measure of the purely quantum
effects in the equations of motion. Nevertheless, as we will be shown below when considering
initial conditions of higher energy, this is not generic. In fact, this is a property
of the nonlinearity of the equations: the effects of a term of order $\hbar^2$ on the
equations of motion are not necessarily of the same order in the solution.

Finally, it is of interest to analyze the time evolution of the terms
$\delta_1 q$ and $\delta_2 q$. Note that both are periodic functions,
with approximately the same period as the classical system $T$,
with an amplitude that increases with time. In fact, $\delta_1 q$
and $\delta_2 q$ follow the same pattern, that is, they have
qualitatively the same form, but with a phase difference of $T/4$
so that when one of them is at a maximum (or at a minimum)
the other one is around zero. In the case of $\delta_1 p$ and $\delta_2 p$,
they are also periodic functions with period $T$, follow the same pattern
and $T/4$ dephased. The main difference between the pattern followed
by $\delta_1 q$ and $\delta_2 q$ with respect to the one followed by 
$\delta_1 p$ and $\delta_2 p$ is that, whereas the formers have just
a critical point between consecutive changes of sign, the latters
oscillate twice (producing three critical points) between two of their zeros.

The net result of all commented effects on the phase-space orbits
can be seen in the lower plot of Fig. \ref{figp010}. Minimum departure
from classical orbit occurs at turning points and when $q$ crosses
the origin. In this plot it is possible to see again that $\delta_1$ and $\delta_2$ follow
qualitatively the
same pattern but, interestingly, they are (almost) not dephased;
the phase differences in position and in momentum compensate each other.
In a more detailed level, it is possible to observe that critical
points of $(\delta_1q^2+\delta_1p^2)$ and $(\delta_2q^2+\delta_2p^2)$
do not exactly coincide in
time: there is a slight delay between them. In addition,
from these plots it can also be inferred that the orbit followed by
the expectation values of quantum states does not coincide
at any point with its classical counterpart, since there is
no time when all corrections vanish: $\delta_1q=\delta_2q=\delta_1p=\delta_2p=0$.

\begin{figure}
\begin{center}
\includegraphics[width=0.5\textwidth]{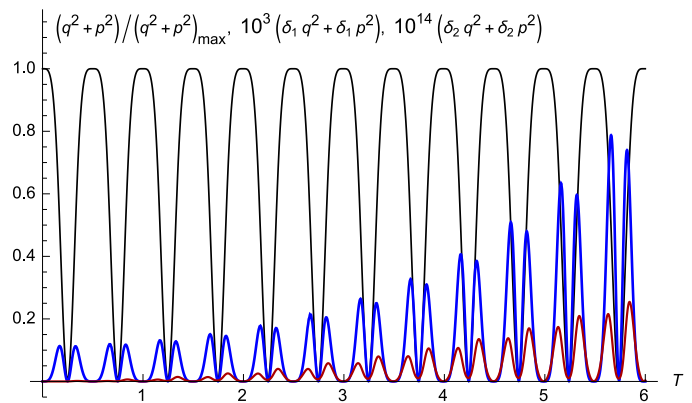}
\end{center}
\caption{In this figure the initial value of the momentum is $p(t_0)=100$. Note
that enhancement factors, by which differences between solutions are multiplied, differ from
the previous case.}\label{figp0100}
\end{figure}
$v/$ Finally, regarding the correspondence principle, it is necessary
to relate the results commented above for the case $p(0)=10$,
with results obtained for larger values of the initial condition
of the momentum. In particular, Fig. \ref{figp0100} shows
the plot equivalent to the last graphic of Fig. \ref{figp010}
for the initial value $p(0)=100$. As already commented above,
the qualitative behavior of the system does not change.
Nonetheless, there are significant modifications in
quantitative aspects that lead us to conclude that the
behavior is more classical.

First of all we notice that the larger the value of $p(0)$,
the longer (in terms of its period)
the system stays stable. This is due to the fact
that the corrections due
to the moments are relatively smaller and take longer to
move the system significantly from its classical trajectory.
More precisely, as can be seen in Fig. \ref{figp0100},
the system has to be evolved during six cycles so that
the departure from the classical trajectory, dominated
by distributional effects, $(\delta_1q^2+\delta_1p^2)$
is of the same order of magnitude as the one obtained
for the previous ($p(0)=10$) case with just two cycles.

In addition, as another important result of this paper,
we note that the relative importance between the two
quantum effects, which can be measured by the quantity
\begin{equation}\label{gamma}
\gamma:=(\delta_2q^2+\delta_2p^2)/(\delta_1q^2+\delta_1p^2),
\end{equation}
is smaller the larger the energy of the system.
That is, from Fig. \ref{figp0100}, we get
$\gamma\approx10^{-11}$ for the case with larger energy
($p(0)=100$), whereas $\gamma\approx10^{-8}$ for
the previous less-energetic case with $p(0)=10$.
The case $p(0)=1$ has also been checked for which,
after a little bit more than half a cycle, the following
values are measured: $(\delta_1q^2+\delta_1p^2)\approx10^{-3}$
and $(\delta_2q^2+\delta_2p^2)\approx10^{-8}$. These results
give $\gamma\approx 10^{-5}$ for the case $p(0)=1$.
This result shows that the quantity $\gamma$ defines
a semiclassical behavior of a system when its value
is small. Nonetheless, when $\gamma$ tends to zero there are
still distributional effects present. This shows that, as commented in
the introduction, the classical limit of a quantum state
is an ensemble of classical trajectories described, in
this context, by its corresponding classical moments.

\section{Conclusions}\label{sec_conclusions}

In this paper the formalism presented in Ref. \cite{Bri14}, to
analyze the evolution of classical and quantum probability
distributions, has been applied to the system of a particle
on a potential.
Due to the kinetic term,
the Hamiltonian of this system is quadratic in the momentum,
and its dependence
on the position is completely encoded in the potential.
The special properties of the harmonic Hamiltonians, which
are defined as those that are at most quadratic on the
basic variables, makes them much easier to be analyzed.
Thus, the study has been divided in two different sectors.
On the one hand, the complete set of harmonic Hamiltonians has been studied;
and, on the other hand, for the anharmonic case an interesting example
has been chosen: the pure quartic oscillator.

By choosing different functional forms of the potential,
three physically different harmonic Hamiltonians can be constructed.
First, the system of a particle moving under a uniform force, which
also includes the free particle when the value of this force
is considered to be zero. Second, the harmonic oscillator with 
a constant frequency $\omega$. And finally the inverse harmonic
oscillator, which can be understood as a harmonic oscillator with
imaginary frequency. For all of them the moments corresponding
to their stationary and dynamical states have been explicitly obtained.
In this framework the stationary states correspond to fix points of
the dynamical system, which is composed by the infinite set of equations
of motions for expectation values and moments. Therefore, in order
to find these stationary moments, the algebraic system obtained
by dropping all time derivatives must be solved. With this procedure,
and contrary to the usual treatment of considering the time-independent
Schr\"odinger equation, the stationary moments can be obtained
without solving any differential equation.

More precisely, regarding the particle under a uniform force, it has been shown that
even if the classical (distributional) case accepts a stationary
state where the particle is at rest at any position and with arbitrary
value of its corresponding (high-order) fluctuations, such a state is
forbidden in the quantum system by the Heisenberg uncertainty principle.
For the harmonic oscillator, the moments corresponding to any stationary
state have been obtained in terms of the frequency of the oscillator
and the energy of the state. These relations are valid for any stationary
state. The only ingredient that is not derived by the present formalism,
and thus one needs to include by hand, are the eigenvalues of the energy.
Finally, it has been proven that
the inverse harmonic oscillator can not have stationary states.

Concerning the pure quartic oscillator, the moments corresponding to
any stationary state have been derived by making use of the above technique.
In this case, the system of equations is not complete and thus it does not fix
the whole set of moments. Hence, apart from the energy of the state,
the fluctuation of the position has been left as a free parameter. Furthermore,
in order to constraint the values of these two parameters,
use has been made of the high-order inequalities which
were derived in \cite{Bri14}. For the particular case of the ground state,
a reasonable assumption is that the Heisenberg uncertainty relation is saturated.
This leads to a tight interval for the value of the ground energy (\ref{interval6}).
It turns out that the exact (numerically computed) value of this energy is not contained in this interval,
but it is quite close. Therefore one can assert that, even if
the exact saturation of the Heisenberg uncertainty relation provides a
good approximation for the ground state of the pure quartic oscillator,
it is not exactly obeyed.

The above analysis shows the practical relevance of the inequalities
that were derived in \cite{Bri14} as a complementary method to extract
physical information from the system. In particular, high-order inequalities
are of relevance because the conditions they provide are stronger
than the ones obtained from lower-order inequalities.

Finally, a numerical computation of the dynamical states corresponding to
the pure quartic oscillator has been performed. To that end, a Gaussian
in the position has been assumed as the initial state. In this setting,
a number of interesting results have been obtained.

First, the validity of the method
has been analyzed. The present formalism is valid as long as the
high-order moments that one drops with the cutoff are small.
The natural tendency of the moments in this system
is to oscillate with a growing amplitude and thus, from certain point on,
this method will not give trustable results. In order to find
the region of validity of the method, on the one hand, different cutoffs
have been considered and the convergence of the solution with the
cutoff order has been studied. On the other hand, the conservation
of the Hamiltonian, as well as the fulfillment of the high-order
inequalities mentioned above, has been monitored during the evolution.
With these control methods at hand, one can estimate when (after
how many cycles) the formalism is not valid anymore.
In particular, this ``validity time'' increases with the value of the
initial classical energy.

Second, the departure of the centroid from its classical point trajectory
has been analyzed, as well as the relative relevance of the
two different quantum effects: the distributional and
the purely quantum effects. It has been shown that, as one would expect,
the former ones, which are also present in the evolution of a
classical probability distribution, are much more relevant
than the latter ones.
Nonetheless, the strength of the purely quantum effects
in the equations of motion is of order $\hbar^2$. Therefore, a change in
the numerical value of the Planck constant would tune the relative relevance
of these effects.

Finally, the correspondence principle has also been verified
in the sense that the larger the classical initial value of the energy is chosen,
the smaller purely quantum effects are measured. In particular,
the smallness of the quantity $\gamma$, as defined in (\ref{gamma}), gives a precise
notion of semiclassicality. In fact the vanishing of $\gamma$
would define a complete classical (distributional) behavior of the system.
Let us stress the fact that this classical behavior is distributional.
In other words, and as commented already throughout the paper, the classical
limit of a quantum state is not a unique orbit on the phase space but, instead,
an ensemble of classical trajectories which are described by a probability distribution
or, in the context of the present formalism, by its classical moments.

\acknowledgments

The author thanks Carlos Barcel\'o, Ra\'ul Carballo-Rubio, I\~naki Garay, Claus Kiefer,
Manuel Kr\"amer, and Hannes Schenck
for discussions and comments. Special thanks to Martin Bojowald for interesting comments
on a previous version of this manuscript. Financial support from the Alexander von
Humboldt Foundation through a postdoctoral fellowship is gratefully acknowledged.
This work is supported in part by Projects IT592-13 of the Basque Government
and FIS2012-34379 of the Spanish Ministry of Economy and Competitiveness.

\end{document}